%% file: main.tex
\documentclass{article}

\usepackage{microtype}
\usepackage{graphicx}
\usepackage[version=4]{mhchem}
\usepackage{subcaption}
\usepackage{booktabs} 

\usepackage[table,xcdraw]{xcolor} 
\usepackage{pifont}
\newcommand{\cmark}{\ding{51}}%
\newcommand{\xmark}{\ding{55}}%
\usepackage{multirow}
\usepackage{nicematrix}
\usepackage[utf8]{inputenc}

\usepackage[accepted]{icml2025}

\usepackage{amsmath}
\usepackage{amssymb}
\usepackage{mathtools}
\usepackage{amsthm}
\usepackage{xcolor}         
\usepackage{placeins}

\definecolor{royalazure}{rgb}{0.0, 0.22, 0.86}
\definecolor{royalblue}{rgb}{0.04,0.33,0.64 }
\definecolor{red}{HTML}{FF2100}
\definecolor{orange}{HTML}{FFA100}
\definecolor{lightgreen}{HTML}{BDE16D}

\usepackage[
    colorlinks,
    citecolor=royalazure,
    filecolor=royalazure,
    linkcolor=royalazure,
    urlcolor=blue
]{hyperref}
\usepackage[normalem]{ulem}
\usepackage{amsmath}
\usepackage{algpseudocode}
\usepackage{amsfonts}
\usepackage{array}
\usepackage{caption}
\usepackage{enumitem}
\usepackage{tabularx} 
\usepackage{booktabs}
\usepackage{colortbl} 
\usepackage{pifont}

\usepackage[capitalize,noabbrev]{cleveref}

\theoremstyle{plain}

\theoremstyle{definition}

\theoremstyle{remark}

\usepackage[textsize=tiny]{todonotes}

\icmltitlerunning{Retro-Rank-In}

\begin{document}

\twocolumn[
\icmltitle{
Retro-Rank-In: \\ 
A Ranking-Based Approach for Inorganic Materials Synthesis Planning
}




\begin{icmlauthorlist}
\icmlauthor{Thorben Prein}{tum,ds,ai,int}
\icmlauthor{Elton Pan}{mit}
\icmlauthor{Sami Haddouti}{tum,ai}
\icmlauthor{Marco Lorenz}{tum,ai}
\icmlauthor{Janik Jehkul}{tum,ai}
\icmlauthor{Tymoteusz Wilk}{tum,ai}
\icmlauthor{Cansu Moran}{tum,ai}
\icmlauthor{Menelaos Panagiotis Fotiadis}{tum,ai}
\icmlauthor{Artur P. Toshev}{tum}
\icmlauthor{Elsa Olivetti}{mit}
\icmlauthor{Jennifer L.M. Rupp}{tum,int}
\end{icmlauthorlist}

\icmlaffiliation{tum}{Technische Universit\"at M\"unchen}
\icmlaffiliation{ai}{TUM.ai}
\icmlaffiliation{int}{TUMint. Energy Research GmbH}
\icmlaffiliation{ds}{Munich Data Science Institute}
\icmlaffiliation{mit}{Massachusetts Institute of Technology}

\icmlcorrespondingauthor{Jennifer L.M. Rupp}{jrupp@tum.de}

\icmlkeywords{Machine Learning, ICML}

\vskip 0.3in
]



\printAffiliationsAndNotice{}

\begin{abstract}

Retrosynthesis strategically plans the synthesis of a chemical target compound from simpler, readily available precursor compounds. This process is critical for synthesizing novel inorganic materials, yet traditional methods in inorganic chemistry continue to rely on trial-and-error experimentation. While emerging machine-learning approaches struggle to generalize to entirely new reactions due to their reliance on known precursors, as they frame retrosynthesis as a multi-label classification task. To address these limitations, we propose \textit{Retro-\mbox{Rank-In}}, a novel framework reformulating the \textbf{Retro}synthesis problem by embedding target and precursor materials into a shared latent space and learning a pairwise \textbf{Rank}er on a bipartite graph of \textbf{In}organic compounds. We evaluate Retro-\mbox{Rank-In's} generalizability on challenging retrosynthesis dataset splits designed to mitigate data duplicates and overlaps. For instance, for \ce{Cr2AlB2}, it correctly predicts the verified precursor pair \ce{CrB} + \ce{Al} despite never seeing them in training, a capability absent in prior work. Extensive experiments show that Retro-Rank-In sets a new state-of-the-art, particularly in out-of-distribution generalization and candidate set ranking, offering a powerful tool for accelerating inorganic material synthesis.  
\end{abstract}

\section{Introduction}
The discovery of inorganic materials underpins a wide array of modern technologies, such as renewable energy and electronics. Recent efforts involving large-scale computational exploration of the materials chemical space \cite{sriram2024open, merchant2023scaling, kim2021solid, zhu2024uncovering} have led to the discovery of millions of potentially stable and synthesizable compounds (\textit{what} to synthesize) \cite{merchant2023scaling, barroso2024open, saal2013materials, zeni2023mattergen}. However, the synthesis of these novel materials remains a critical bottleneck (\textit{how} to synthesize) \cite{karpovich2023interpretable, mahbub2020text}. Unlike inorganic materials, organic molecules exist as discrete, individual structures, which allow their synthesis to be broken down into multiple steps, each with smaller building blocks through a well-understood sequence of reactions -- a process called \textit{retrosynthesis}. In contrast, inorganic materials adopt a periodic structure 3D arrangement of atoms. This periodicity renders the retrosynthetic strategy known in organic synthesis inapplicable to inorganic materials. The synthesis of inorganic materials largely remains a one-step process, where a \textit{set} of precursors undergo a reaction to form a desired target compound. This complex process has no general, unifying theory \cite{kononova2019text}, and thus heavily relies on trial-and-error experimentation of precursor materials. Furthermore, the exponential 
scaling of compute needed for simulation impedes physical modeling of the underlying physical phenomena, e.g., thermodynamics and kinetics at the atomic scale \cite{bianchini2020interplay}. 

\begin{figure}[!t]
    \centering
    \includegraphics[width=1\linewidth]{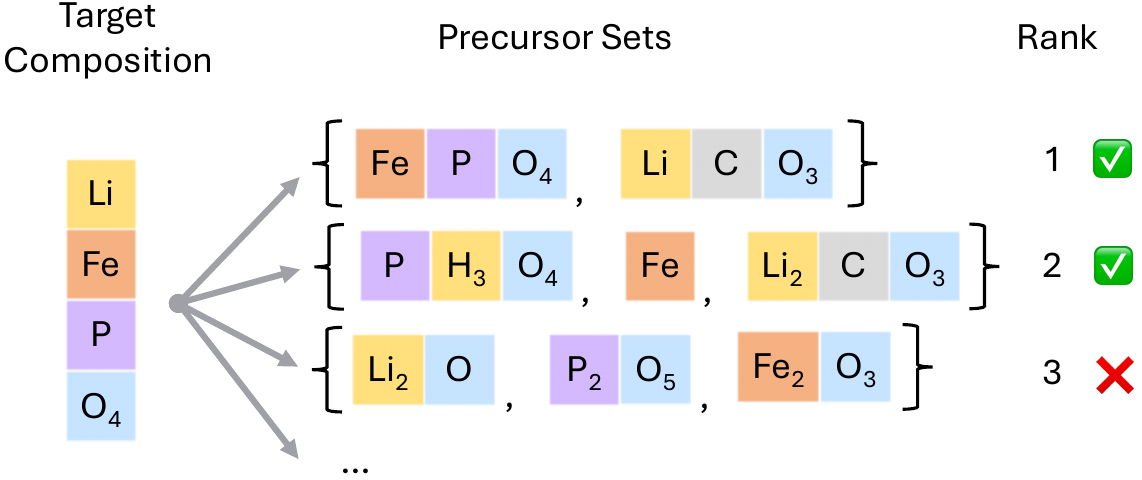}
    \caption{
    \textbf{Retrosynthesis problem.} 
    Identifying the optimal precursor set for a given target material can be treated as a ranking problem. We use the binary classification probabilities of each set to determine its rank. Checkmarks indicate whether a ranked set corresponds to an experimentally verified synthesis.
    }
    \label{fig:1}
\end{figure}

This presents a compelling opportunity for machine learning (ML) approaches to bridge the knowledge gap by learning directly from synthesis data. In particular, precursor recommendation stands out as a key task in inorganic materials synthesis \citep{miura2021observing, bianchini2020interplay}. For a reaction $A + B \rightarrow C$, the task is to recommend a set of precursors $\{A, B\}$ given target $C$. Early work in the field utilized a text-conditioned conditional variational autoencoder to generate synthesis precursors for novel materials \cite{kim2020inorganic}. ElemwiseRetro \cite{kim2022element} employs domain heuristics and a classifier for template completions. More recently, studies have leveraged language models to uncover and analyze relationships between target materials and their precursors \cite{kim2024large}. An orthogonal approach trains a reaction template retriever by learning representations of target materials using a masked precursor completion task. These learned representations are then used to retrieve records of known syntheses of materials similar to the target material, which achieves strong performance in precursor recommendation \cite{he2023precursor}.

Notably, the most recent work Retrieval-Retro \cite{noh2024retrieval} employs two retrievers, the first identifying reference materials
sharing similar precursors with the target material, while the second retriever suggests precursors based on formation energies. Specifically, this approach uses self-attention and cross-attention for target-reference material comparison and predicts precursors via a multi-label classifier.
This framework effectively unifies both a data-driven and a domain-informed approach for inorganic retrosynthesis. 

However, existing ML approaches face significant limitations, as summarized in \cref{tab:related_work_comparison}. Most notably, they lack the ability to incorporate new precursors, a critical aspect of experimental workflows in laboratories when searching for novel precursors and discovering new compounds \cite{mcdermott2023assessing, szymanski2024computationally}. For instance, Retrieval-Retro \cite{noh2024retrieval} cannot recommend precursors outside its training set, as they are represented through one-hot encoding in its multi-label classification output layer (\cref{fig:2}, a.). This design restricts the model to recombining existing precursors into new combinations rather than enabling predictions involving entirely novel precursors that have not been seen during training, thereby limiting its applicability in a material discovery setting. Furthermore, prior methods struggle to effectively incorporate broader chemical knowledge. Retrieval-Retro utilizes a Neural Reaction Energy (NRE) retriever trained to predict formation enthalpy using the Materials Project DFT database of approximately 80,000 computed compounds \cite{jain2013commentary}, but this approach does not fully exploit domain-specific data. Another limitation in extrapolation capabilities arises from the embedding design in previous approaches. Specifically, these methods embed precursor and target materials in disjoint spaces, which hinders their ability to generalize effectively. 

\begin{table}[tb]
\centering
\caption{\textbf{Comparison of precursor planning methods.}  ElemwiseRetro (template-based), Synthesis Similarity \& Retrieval-Retro (retrieval-based), and our ranking-based approach compared for model capabilities.}
\label{tab:related_work_comparison}
\resizebox{\linewidth}{!}{
\begin{NiceTabular}
{lc@{\hspace{0.6cm}}cc@{\hspace{0.6cm}}c}
\toprule
\textbf{Model} & \textbf{\shortstack{Discover \\ new precursors}} & \textbf{\shortstack{Chemical \\  domain knowledge}} & \textbf{\shortstack{Extrapolation \\ to new systems}} \\
\midrule
\arrayrulecolor{white}
ElemwiseRetro
 &  & \cellcolor{red!40} & \cellcolor{orange!35} \\
 \cite{kim2022element} & \multirow{-2}{*}{\textcolor{red}{\xmark }} & \multirow{-2}{*}{\cellcolor{red!40}Low} & \multirow{-2}{*}{\cellcolor{orange!35}Medium} &\\
\hline
Synthesis Similarity  &  & \cellcolor{red!40} & \cellcolor{red!40} \\
\cite{he2023precursor} & \multirow{-2}{*}{\textcolor{red}{\xmark }} & \multirow{-2}{*}{\cellcolor{red!40}Low} & \multirow{-2}{*}{\cellcolor{red!40}Low} &\\
\hline
Retrieval-Retro & & \cellcolor{red!40} & \cellcolor{orange!35} \\
\cite{noh2024retrieval} & \multirow{-2}{*}{\textcolor{red}{\textcolor{red}{\xmark}}} & \multirow{-2}{*}{\cellcolor{red!40}Low} & \multirow{-2}{*}{\cellcolor{orange!35}Medium} &\\
\hline
\textbf{Retro-\mbox{Rank-In}} & & \cellcolor{orange!35} & \cellcolor{green!20} \\
\textbf{(Ours)} & \multirow{-2}{*}{\textcolor{blue}{\cmark}} & \multirow{-2}{*}{\cellcolor{orange!35}Medium} & \multirow{-2}{*}{\cellcolor{green!20}High} &\\
\arrayrulecolor{black}
\bottomrule
\end{NiceTabular}}
\normalsize 
\end{table}

To address these gaps, we propose \textbf{Retro-Rank-In}, a unified framework for identifying and ranking precursor sets (\cref{fig:2}). Retro-\mbox{Rank-In} consists of two core components: a composition-level transformer-based materials encoder, which generates chemically meaningful representations of both target materials and precursors, and a Ranker that evaluates chemical compatibility between the target material and precursor candidates. The Ranker is specifically trained to predict the likelihood that a target material and a precursor candidate can co-occur in viable synthetic routes.

Our key contributions are as follows:  
\begin{itemize}  
\item \textit{Increased flexibility}: During inference, Retro-\mbox{Rank-In} enables the selection of new precursors not seen during training. This is crucial for exploring novel compounds as it allows the incorporation of a larger chemical space into the search for new synthesis recipes \cite{mcdermott2023assessing}.  
\item \textit{Incorporation of broad chemical knowledge}: We leverage large-scale pretrained material embeddings to integrate implicit domain knowledge of formation enthalpies and related material properties.  
\item \textit{Joint embedding space}: By training a pairwise ranking model, we embed both precursors and target materials within a unified embedding space, thereby enhancing the model's generalization capabilities. 
\item \textit{Analysis of sequential models}: We compare our method against autoregressive generation approaches, demonstrating that our framework provides a more robust and accurate alternative, particularly for tasks requiring the simultaneous evaluation of multiple precursors instead of sequential modeling.  
\end{itemize}


\section{Preliminaries} \label{sec:preliminaries}

\begin{figure*}[!htb]
    \centering
    \includegraphics[width=1\textwidth]{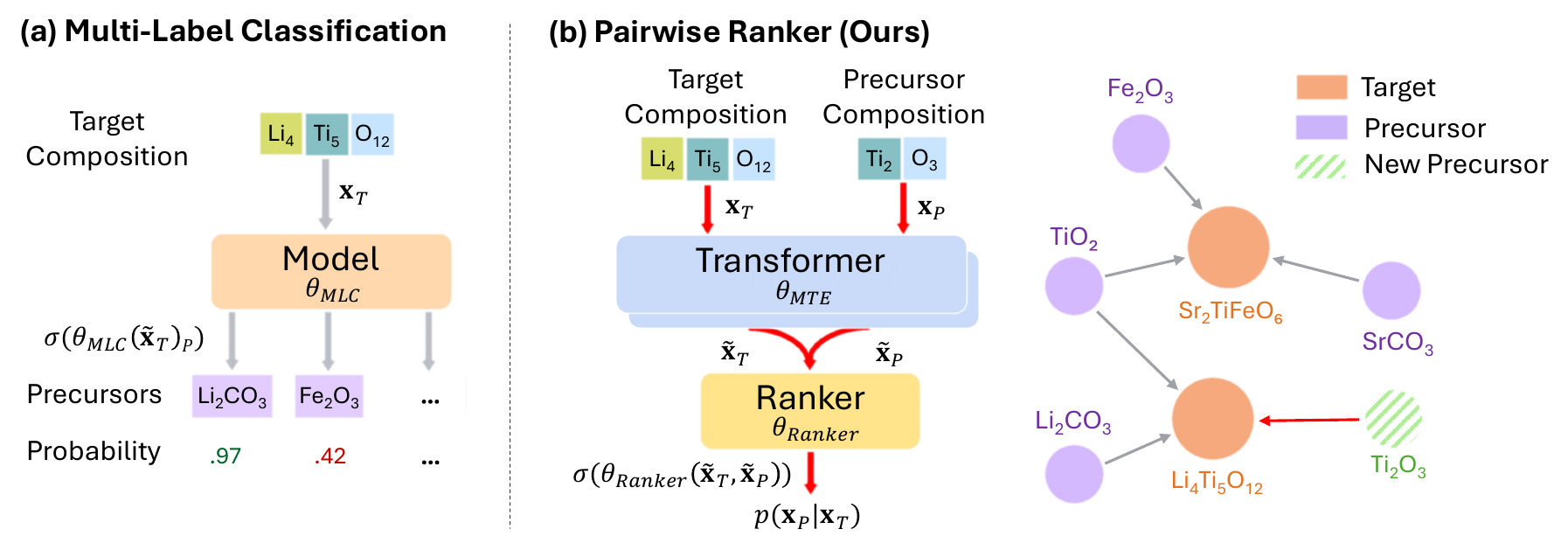} 
    \caption{\textbf{Learning paradigms for inorganic retrosynthesis} 
    (a) Multi-label classification-based approaches, which constitute current state-of-the-art models \cite{noh2024retrieval}, inherently predict known precursors $P$ from a fixed candidate set. (b) Our approach (Retro-\mbox{Rank-In}) overcomes this limitation by embedding both precursor and target materials into a shared latent space and predicting their chemical compatibility in synthetic routes. This enables extrapolation beyond known precursors, allowing the model to propose novel synthesis pathways for unseen materials. The red links highlight an exemplary case for link prediction between a target and a precursor.}
    \label{fig:2}
\end{figure*}

\paragraph{Retrosynthesis.} During a forward synthesis process in inorganic chemistry, a set of precursors $P_1, \dots, P_m$ are mixed and heated to form a target material $T$. The inverse problem, namely retrosynthesis, devises a combination of precursors as a set that reacts to form a desired target material. Retrosynthesis involves working backward from a target compound (e.g., \ce{Li7La3Zr2O12}) to deduce a set of simpler precursor compounds (e.g., $\{$\ce{LiOH}, \ce{La2O3}, \ce{ZrO2}$\}$) that can feasibly synthesize the desired product. However, this is an under-determined problem as there are many possible sets of valid precursors, which, under the right reaction conditions, might form the target compound. Moreover, the feasibility of a given precursor set can be quantified by considering factors such as the required synthesis pressure and temperature, the cost of precursor materials, and the yield of the process. Inspired by the fact that some precursor combinations are more advantageous than others, we formulate the learning problem as a ranking task over precursor sets.

\paragraph{Learning problem.} 

Building upon previous work \citep{he2023precursor, noh2024retrieval}, our objective is to predict a ranked list of precursor sets, denoted as 
\begin{equation}
    (\mathbf{S}_1, \mathbf{S}_2, \ldots, \mathbf{S}_K).
\end{equation} 
Each precursor set $\mathbf{S} = \{ P_1, P_2, \ldots, P_m \}$
consists of \(m\) individual precursor materials, where the number of precursors \(m\) can vary for each set.
The ranking indicates the predicted likelihood of each precursor set forming the target material. Historically reported synthesis routes from the scientific literature are considered correct predictions.

While prior work focuses on learning a multi-label classifier \(\theta_{MLC}\) over a predefined set of precursors/classes \citep{he2023precursor, noh2024retrieval}, we redefine the problem to learn a pairwise ranker $\theta_{\text{Ranker}}$
of a precursor material $P$ conditioned on target $T$, see \cref{fig:2}. 
Thus, our reformulation of the learning problem enables inference on entirely novel precursors and precursor sets, a capability that has not been achieved by previous methods. In addition, we know that datasets in chemistry are highly imbalanced, with a large number of possible precursors and only a few positive labels. Our pairwise scoring approach allows for custom sampling strategies, including negative sampling, to improve balance.

\paragraph{Compositional representation.}
For a given target material $T$, we represent its elemental composition as a vector \(\mathbf{x}_T = (x_1, x_2, \dots, x_d)\), where each \(x_i\) corresponds to the fraction of element \(i\) in the compound, and $d$ is the count of all considered chemical elements. For example, titanium dioxide (\ce{TiO2}) can be expressed as a composition vector where titanium (Ti, atomic number 22) and oxygen (O, atomic number 8) contribute respective fractions \(x_{22} = \frac{1}{3}\) and \(x_{8} = \frac{2}{3}\).

\section{Retro-Rank-In} \label{sec:architecture}

\subsection{Embedding model} \label{subsec:embedding_model}

We use a multi-task pretrained transformer-based encoder $\theta_{\text{MTE}}$, 
adapted from \citet{prein2023mtencoder} and \citet{wang2021compositionally}, to map each composition 
\(\mathbf{x}\) to an $h$-dimensional latent representation \(\tilde{\mathbf{x}} = \theta_{\text{MTE}}(\mathbf{x})\), with 
\(\tilde{\mathbf{x}} \in \mathbb{R}^{h}\).

\paragraph{Input sequence construction.}
First, each element \(i\) with a non-zero component is mapped to a learned elemental embedding \(\mathbf{e}_i\). During pretraining, the model is initialized with mat2vec elemental embeddings \cite{murdock2020domain}. To account for the continuous stoichiometric fraction \(x_i\), we apply a sinusoidal fractional embedding \(\mathbf{f}_i\), inspired by positional encodings in transformers \citep{vaswani2017attention}. We then sum the elemental embeddings and their fractional encodings to form a single per-element embedding:
\begin{equation}
    \mathbf{z}_i = \mathbf{e}_i + \mathbf{f}_i.
\end{equation}
To achieve rich compound representations, we learn a special \texttt{[CPD]} token \(\mathbf{t}\) that is prepended to the sequence to serve 
as a global compound-level representation. 
Thus, the final input sequence \(\mathbf{s}\) becomes:
\begin{equation}
    \mathbf{s} = \bigl[ \mathbf{t}, \mathbf{z}_1, \mathbf{z}_2, \dots, \mathbf{z}_k \bigr],
\end{equation}
where \(k\) is the number of distinct elements in the composition. 
We pass \(\mathbf{s}\) through three transformer encoder blocks:
\begin{equation}
    \tilde{\mathbf{x}} = \theta_{MTE}\bigl(\mathbf{s}\bigr),
\end{equation}
which use self-attention to contextualize each element’s embedding. 

\paragraph{Multi-task pretraining.} To encourage broader generalization, we pretrain the encoder on the Alexandria database \cite{hoffmann2023transfer} of over two million unique compositions using a multi-task objective, as described in \cref{appendix_mte}. This includes (i) \textit{masked element prediction}, where randomly masked elements are reconstructed, (ii) \textit{DFT-property regression}, predicting 10 computed properties (e.g., formation enthalpy, stress tensor component, band gap) by feeding the final hidden state of \texttt{[CPD]} into regression heads, and (iii) \textit{space-group classification}, determining the space group of $\mathbf{x}$ from the same special token output.

\subsection{Ranker} \label{subsec:ranker}
We introduce a binary classifier $\mathcal{B}: \mathbb{R}^h \times \mathbb{R}^h \rightarrow \mathbb{P}$ to evaluate precursor-target pair relevance. Given a target representation $\tilde{\mathbf{x}}_T \in \mathbb{R}^h$ and precursor representation $\tilde{\mathbf{x}}_P \in \mathbb{R}^h$, we compute the probability $p \in \mathbb{P}$ of the precursor forming a valid synthetic path to the target. The training dataset consists of balanced positive and negative pairs, effectively forming a bipartite edge prediction problem akin to recommender systems \cite{gao2022graph}.

During inference, for a given target $\mathbf{x}_T$, we compute $p(\mathbf{x}_P | \mathbf{x}_T)=\theta_{\text{Ranker}}(\tilde{\mathbf{x}}_T,\tilde{\mathbf{x}}_P)$ for each precursor in the dataset. The top-$K$ precursors, determined by probability ranking, are combined into valid sets satisfying elemental completeness and predefined cardinality constraints \cite{noh2024retrieval}. Assuming precursor independence, we compute the joint probability of each set $S$ as:
\begin{equation}
p_S = \prod_{i=1}^{m} p_i,
\end{equation}
where $m$ denotes the number of precursors in the set. We evaluate the final ranked precursor sets, ordered by $p_S$, using Top-$K$ metrics.

\subsection{Model training}
We collect target--precursor pairs \((\mathbf{x}_T, \mathbf{x}_P)\) from known synthesis routes, and label them \(y \in \{0,1\}\). To balance classes, we sample positive ($y=1$) and negative ($y=0$) pairs with equal probability. 
Each composition \(\mathbf{x}\) is encoded by \(\theta_{\text{MTE}}\), producing \(\tilde{\mathbf{x}}\). For a target--precursor pair \((\mathbf{x}_T, \mathbf{x}_P)\), we concatenate their embeddings \(\bigl[\tilde{\mathbf{x}}_T \,\|\, \tilde{\mathbf{x}}_P\bigr]\) and pass them to the ranker \(\theta_{\text{Ranker}}\) (\cref{fig:2}, b.) to predict the precursor probability \(p(\mathbf{x}_P | \mathbf{x}_T)\).
We train using binary cross-entropy:  
\begin{equation}
     \mathcal{L}_{\text{BCE}} = -\frac{1}{|\mathcal{D}|}\sum_{i\in \mathcal{D}} 
    \Bigl[y_i \log p_i + (1 - y_i) \log(1 - p_i)\Bigr],
\end{equation}
where \(p_i = p(\mathbf{x}_P | \mathbf{x}_T)\). 
At test time, given a target composition \(\mathbf{x}_T\), we compute \(p(\mathbf{x}_P | \mathbf{x}_T)\) for each candidate precursor \(\mathbf{x}_P\) and select the highest-ranked precursors for further evaluation (see \cref{appendix:eval}).

\section{Experiments} \label{sec:experiments}

We begin by detailing our experimental setup, datasets, evaluation protocols, and baseline comparisons. We then present empirical results, followed by an in-depth analysis of our approach, including extensive ablation studies.

\subsection{Setup} \label{subsec:setup}

\paragraph{Datasets.}
\label{Dataset_}

We build upon the dataset introduced by \citet{kononova2019text}, which has been widely used in recent studies \cite{noh2024retrieval, he2023precursor}. Curated from the literature using paragraph- and phrase-level NLP classifiers, this dataset captures solid-state reactions, including byproducts, targets, and precursors, and comprises 33,343 synthesis recipes extracted from published sources.

However, this dataset has some incomplete or ambiguous entries. To ensure data quality, we apply several preprocessing steps to validate chemical formulas. Specifically, we exclude entries containing variables such as \textit{b} and \textit{c} or other symbolic placeholders, retaining only those with explicitly defined stoichiometries and valid element symbols from the periodic table. Additionally, we enforce a constraint requiring that all elements in the target material -- except for C, O, H, and N -- must also be present in the precursor materials. A viable assumption to make is that new elements cannot form during solid-state reactions.

After preprocessing, the dataset consists of 18,804 entries, of which 9,255 are unique, i.e., we exclude permutations of precursor sets. Consequently, over half of the dataset consists of duplicate entries.

Previous studies \cite{noh2024retrieval, he2023precursor, kim2022element} have employed a year-based data split to construct a materials discovery setting, using data reported up to and including 2014 for training and validation, while reports after 2014 serve as the test set. However, we recognize that a high number of duplicated synthesis recipes can inflate a model's performance metric in predicting synthesis routes for novel materials. This prevalence of duplicate entries aligns with observations in related fields, such as organic retrosynthetic planning \cite{bradshaw2025challenging}, where repeated recipes are frequently encountered due to recurring synthesis reports. To address these limitations, we propose augmenting the existing year-based split with two additional evaluation settings, resulting in three distinct datasets that present a more challenging setup for assessing extrapolation. More details on the datasets can be found in \cref{Appendix:Dataset}. 

\begin{itemize} 
\item \textbf{Complete Reaction Archive (CRA)}: Includes all entries, retaining duplicate precursor-target combinations, as conducted in previous works \citep{noh2024retrieval, he2023precursor, kim2022element}. 

\item \textbf{Distinct Reactions (DR)}: This dataset focuses exclusively on unique precursor-target combinations, represented as $\{\mathbf{x}_{T}, \mathbf{x}_{P_1}, \ldots, \mathbf{x}_{P_m}\}$. Duplicate entries are removed, ensuring that each reaction is represented only once. This split emphasizes the model's ability to learn distinct chemical pathways. 

\item \textbf{Novel Material Systems (NMS)}: Ensures that no material system -- defined by the set of elements in the target material -- overlaps between the training and test sets. For instance, $\text{Fe}_a\text{P}_b$ samples (where $a, b > 0$) appear either only in the train/validation split or in the test split. This setting enables the evaluation of the model's ability to extrapolate to entirely new systems.
\end{itemize}

\paragraph{Evaluation.} 

Following recent works \citep{noh2024retrieval, he2023precursor, kim2022element}, we employ Top-K exact match accuracy to assess the performance of our binary classifier $\mathcal{B}$ in the inorganic retrosynthesis task. Additional details are provided in \cref{appendix:eval}.

Let the ground-truth precursor set for a target composition \( \mathbf{x}_T \) be denoted as \(\mathbf{S}_{\text{true}}\), with its length defined as \( m = |\mathbf{S}_{\text{true}}|\). Based on the probabilities $p_i$ predicted by the model, we construct precursor sets as described in \cref{subsec:ranker}. A valid set is defined by encompassing all elements of the target compound. These sets are then sorted in descending order of their joint probability. For the Top-K exact match accuracy, we select the Top-K subset of the sorted sets and compare the ground-truth precursor set \(\mathbf{S}_{\text{true}}\) against each set. If a match is found, the corresponding set is assigned a score of 1; otherwise, it is scored as 0. This process is repeated for all target compositions in the test dataset, and the scores are averaged to compute the overall score.

\paragraph{Baselines.} We evaluate our approach against three baseline methods for inorganic synthesis planning, as proposed in prior literature \citep{noh2024retrieval, he2023precursor, kim2022element}. 
These methods formulate precursor prediction as multi-label classification, outputting a vector of dimension $N$, where $N$ represents the number of unique precursors in the dataset. 
\citet{he2023precursor} introduced a masked precursor completion (MPC) task, where attention layers are employed to contextualize the representations of precursors and target materials. The model utilizes these representations to reconstruct the precursor materials. Subsequently, these learned representations are leveraged to search a knowledge base of known target materials, facilitating the transfer of precursors from known materials to novel target materials.
\citet{noh2024retrieval} expands on this by adding a neural reaction energy (NRE) retriever to the MPC task. The NRE module employs a pretrained formation enthalpy predictor to retrieve energetically favorable candidate precursors from a knowledge base. The combined model, namely Retrieval-Retro \citep{noh2024retrieval}, then learns to use information from both modules jointly to predict precursors via a multi-label classification objective. 

The approach proposed by \citet{kim2022element} utilizes a heuristic-based method integrated with a classification task through a multilayer perceptron. It extracts source elements from the input target material composition and predicts corresponding templates from a set of 60 precursor templates. The concatenation of the source element and its correspondingly predicted template forms the precursor compound.

Furthermore, we incorporate three composition-based material representation strategies to predict precursor occurrence via multi-label classification. First, we use a sparse composition approach encoding each element’s fraction in a dedicated input dimension. Second, we apply CrabNet \citep{wang2021compositionally}, a transformer encoder model designed to contextualize elemental embeddings. Finally, we test MTEncoder (see \cref{subsec:embedding_model}), which uses a similar transformer-based architecture but benefits from extensive pretraining on large-scale materials data. In all cases, we pair these representations with feedforward layers that output probabilities for each potential precursor via multi-label classification. 



\begin{table*}[htb!]
\centering
\small 
\caption{\textbf{Performance comparison} Different models were evaluated across three datasets: (a) Complete Reaction Archive, (b) Distinct Reactions, and (c) Novel Material Systems. Bold values indicate the best performance and underline the second best. All scores are reported as averages over five runs, with standard deviations in parentheses.}
\label{table1:performance_comparison}
\resizebox{\linewidth}{!}{
\begin{tabular}
{lcccccccccccccc}
\toprule
 & \multicolumn{4}{c}{\textbf{(a) Complete Reaction Archive}} & \multicolumn{6}{c}{\textbf{(b) Distinct Reactions}} & \multicolumn{4}{c}{\textbf{(c) Novel Materials Systems}} \\
\cmidrule(lr){2-5} \cmidrule(lr){7-10} \cmidrule(l){12-15}
\textbf{Model}      & \multicolumn{4}{c}{\textbf{Top-K Accuracy $\uparrow$}} & \multicolumn{6}{c}{\textbf{Top-K Accuracy $\uparrow$}} & \multicolumn{4}{c}{\textbf{Top-K Accuracy $\uparrow$}} \\
\cmidrule(lr){2-5} \cmidrule(lr){7-10} \cmidrule(l){12-15}
      & Top-1 & Top-3 & Top-5 & Top-10 & & 
      Top-1 & Top-3 & Top-5 & Top-10 & &
      Top-1 & Top-3 & Top-5 & Top-10 \\
\midrule

Composition & 64.72 & 77.68 & 81.17 & 83.44 & & 42.31 & 55.35 & 59.36 & 62.86 & & 43.57 &  55.26 &  58.27 & 60.01 \\
& \scriptsize (0.18) & \scriptsize (0.37) &\scriptsize (0.45) & \scriptsize (0.42) & & \scriptsize (1.15) & \scriptsize (1.00) &\scriptsize(0.95) & \scriptsize (0.80) & & \scriptsize ( 0.37) & \scriptsize ( 0.58) &\scriptsize (0.51) & \scriptsize (0.57) \\



ElemwiseRetro & 64.56 & 70.00 & 71.27 & 72.82 & & 48.72 & 55.19 & 57.30 & 59.08 & & 46.70 & 52.90 & 54.27 & 56.66 \\
\cite{kim2022element} & \scriptsize (0.19) & \scriptsize (0.12) &\scriptsize (0.09) & \scriptsize (0.06) & & \scriptsize (1.15) & \scriptsize (0.32) &\scriptsize (0.21) & \scriptsize (0.22) & & \scriptsize (1.55) & \scriptsize (0.05) &\scriptsize (0.17) & \scriptsize (0.23) \\
SynthesisSimilarity  & 62.25 & 73.19 & 76.03 & 78.51 & & 40.40 & 53.48 & 57.61 & 61.24 & & 36.67 & 47.60 & 50.97 & 53.42\\
\cite{he2023precursor}& \scriptsize (0.75) & \scriptsize (0.55) &\scriptsize (0.43) & \scriptsize (0.30) & & \scriptsize (0.49) & \scriptsize (0.31) &\scriptsize (0.53) & \scriptsize (0.54) & & \scriptsize (0.40) & \scriptsize (0.80) &\scriptsize (1.01) & \scriptsize (0.94) \\

CrabNet & \underline{66.66} & 79.73 &82.98 & 85.24 & & 48.68 & 63.30 &  67.26 & 69.70 & & \underline{48.54} & 59.93 &  62.44 & 63.70 \\
\cite{wang2021compositionally} & \scriptsize (0.43) & \scriptsize (0.16) &\scriptsize (0.37) & \scriptsize (0.22) & & \scriptsize (0.51) & \scriptsize (0.70) &\scriptsize (0.58) & \scriptsize (0.66) & & \scriptsize (0.47) & \scriptsize (0.32) &\scriptsize (0.30) & \scriptsize (0.32) \\

Retrieval-Retro  & 66.22 & 77.45 & 81.01 & 84.28 & & \textbf{49.31} & 62.70 & 67.30 & \underline{71.25} & & 48.05 & 60.77 & 64.26 & \underline{67.68}  \\
\cite{noh2024retrieval}& \scriptsize (0.61) & \scriptsize (0.27) &\scriptsize (0.35) & \scriptsize (0.43) && \scriptsize (0.62) & \scriptsize (0.36) &\scriptsize (0.83) & \scriptsize (0.57) & & \scriptsize (0.58) & \scriptsize (1.26) &\scriptsize (1.18) & \scriptsize (1.29)  \\

MTEncoder& \textbf{67.04}  & \underline{80.53} & \underline{83.89} & \underline{86.10} & & \underline{49.01} & \underline{64.59} & \underline{68.78} & 71.24 & & \textbf{49.35} & \underline{61.98} &  \underline{64.74} & 65.94 \\
& \scriptsize (1.77) & \scriptsize (2.47) &\scriptsize (2.40) & \scriptsize (1.69) & & \scriptsize (0.54) & \scriptsize (0.29) &\scriptsize (0.397) & \scriptsize (0.51) & & \scriptsize (0.40) & \scriptsize (0.54) &\scriptsize (0.59) & \scriptsize (0.45) \\


\midrule
\textbf{Retro-Rank-In (Ours)} & 66.55  & \textbf{80.57} &\textbf{ 85.89} & \textbf{89.85} & & 48.93 & \textbf{65.45} & \textbf{72.51} & \textbf{78.48} & & 47.36 & \textbf{63.27} & \textbf{69.92} & \textbf{76.20} \\
& \scriptsize (0.43) & \scriptsize (0.90) &\scriptsize (0.61) & \scriptsize (0.81) & & \scriptsize (0.50) & \scriptsize (0.31) &\scriptsize (0.69) & \scriptsize (0.82) & & \scriptsize (1.05) & \scriptsize (1.95) &\scriptsize (1.74) & \scriptsize (1.86) \\
\bottomrule
\end{tabular}}
\end{table*}

\subsection{Results}

\cref{table1:performance_comparison} highlights key findings that both align with and extend prior research. Consistent with \citet{noh2024retrieval}, models that explicitly capture interactions between elements (e.g., MTEncoder, CrabNet) outperform simpler composition-based baselines. Notably, MTEncoder benefits from its domain-informed pretraining, which enhances material representations by transferring knowledge from compound properties related to synthesis, such as spacegroup and formation enthalpy, leading to improved performance. All methods achieve high accuracy in the interpolation-focused CRA dataset (\cref{table1:performance_comparison}, a.), 
obtaining strong Top-1 exact match accuracy.
We observe that while methods perform similarly at the Top-1 level, they do show a greater decline in accuracy compared to Retro-\mbox{Rank-In} at higher Top-K settings. We attribute this decline to the imbalanced multi-label training objective, which leads to a skewed probability distribution, many predictions are concentrated near 0 or 1, as shown in \cref{fig:rr_ours_prob_dist}, making it less suitable for the ranking task. Additionally, we see only limited benefit of the retrieval-based augmentation of target material information for methods \citet{noh2024retrieval, he2023precursor}.

To investigate how model performance is influenced by predictions in more challenging extrapolative settings, we curate two new datasets, one of deduplicated, distinct reactions (DR) and another of novel material systems (NMS), are introduced to evaluate model performance in this setting where no material system is shared between the training and test sets.

\paragraph{Distinct reactions.} In the distinct reactions setting (\cref{table1:performance_comparison}, b.), performance declines (compared to the CRA setting) across all methods due to the exclusion of overlapping reactions between training and test data, underscoring the inherent difficulty of true extrapolation beyond memorized training examples. Notably, while the Top-1 gap narrows and Retrieval-Retro surpasses our approach in this metric, Retro-\mbox{Rank-In} maintains high Top-K exact match accuracy across the Top-3, Top-5, and Top-10 evaluations, widening the gap to other approaches. We hypothesize that Retro-\mbox{Rank-In}’s advantage stems from its ability to generate a more smoothly distributed ranking of candidates, leading to better-calibrated scores for non-trivial precursor candidates. We show this in \cref{fig:rr_ours_prob_dist}, where Retro-\mbox{Rank-In} predicts a more diverse set of precursors compared to the next best method. Additionally, we find that MTEncoder achieves competitive performance, highlighting the significance of pretrained and domain-informed embeddings. However, its performance declines for larger values of \( K \), likely due to the challenges associated with the multi-label classification setup. This comparison highlights the robustness of Retro-\mbox{Rank-In} and its ability to generalize effectively even when other methods struggle due to increased deduplication, suggesting that performance gains reported in previous works may have been inflated by the presence of near-duplicate entries.

\paragraph{Novel materials systems.} However, the train-test split in the DR setting still contains compositions that are similar (e.g., \ce{Li5La3Ta2O12} in training vs. \ce{LiLaTa2O7} in testing). As such, we evaluate the models on the most difficult setting (NMS) (\cref{table1:performance_comparison}, c.), which shows a further widening of the train-test gap. All models face greater difficulty in this extrapolative setting, reflecting the complexity of predicting synthesis routes when training and testing data diverge more substantially. In this setting, the domain-informed approaches, Retrieval-Retro and MTEncoder, achieve strong performance for Top-1 set prediction. Notably, Retro-\mbox{Rank-In} performs competitively with Retrieval-Retro on the Top-1 metric.

\begin{figure}[!t]
    \centering
    \includegraphics[width=1\linewidth]{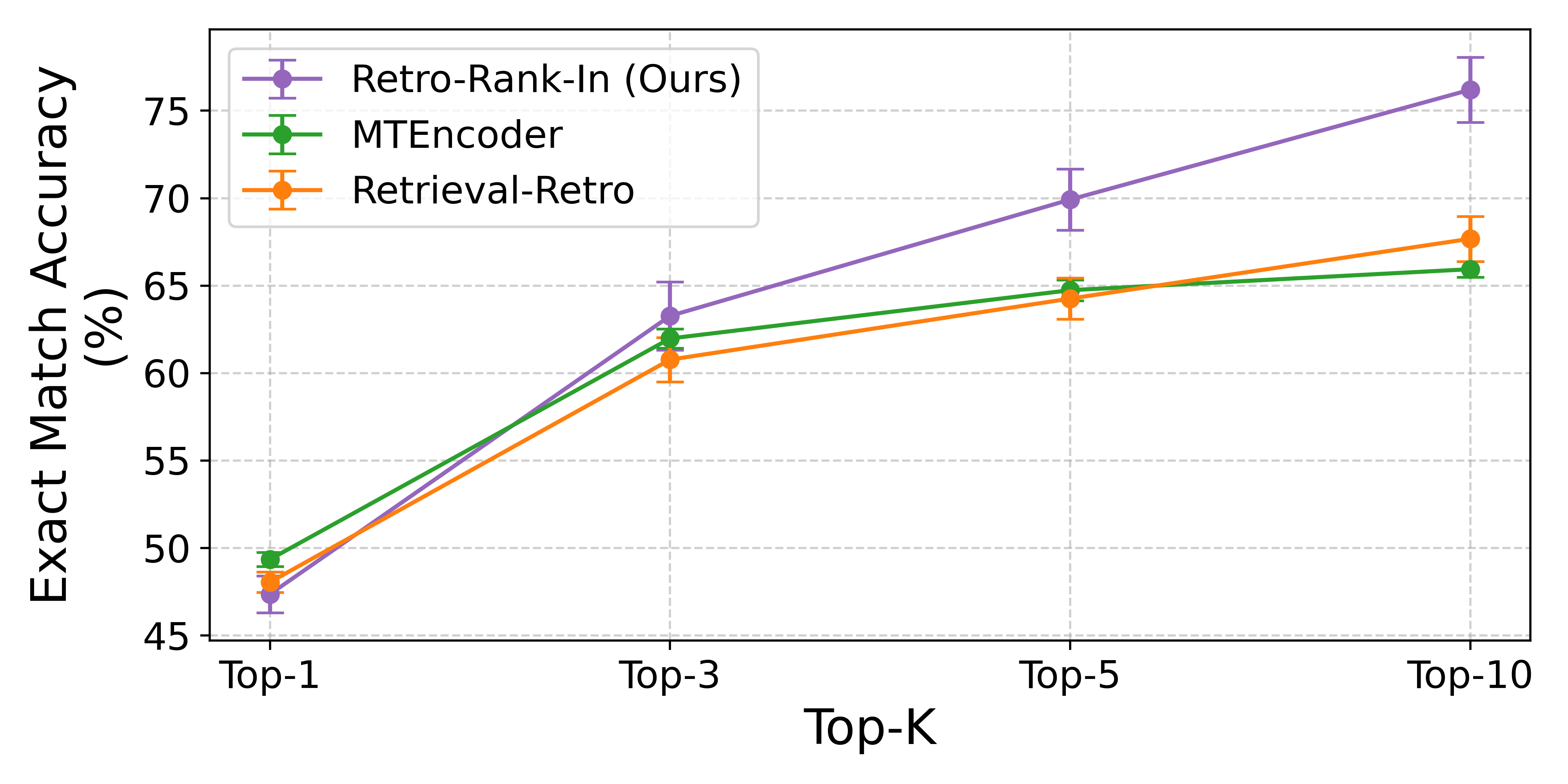}
    \caption{
    \textbf{Comparison of Top-K accuracy.} 
    Comparison of Retrieval-Retro and Retro-\mbox{Rank-In} on the Novel Materials Systems dataset (c). We see the performance gap between both approaches widening, especially for larger K.}
    \label{fig:3}
\end{figure}

\paragraph{Diversity at no cost of performance.} Notably, Retro-\mbox{Rank-In} outperforms the next best models (MTEnocer and Retrieval-Retro) as more precursor sets (Top-$K$) are considered (\cref{fig:3}). Interestingly, the performance gap between the two models widens from 2.5$\%$ for $K=3$ to 8.5$\%$ for $K=10$. This is significant, as this shows that Retro-\mbox{Rank-In} is capable of generating valid precursor sets even at high $K$ values. In contrast, this effect is less pronounced in the baseline methods. We hypothesize that this bifurcation of performance (\cref{fig:3}) at higher values of $K$ is a result of a higher diversity of valid precursor predicted precursors as previously shown in \cref{fig:plot9}. Based on domain knowledge, this result is compelling, as the target-precursor relationship is inherently \textit{one-to-many}, i.e., the same target can be synthesized with multiple possible precursor sets. As such, the high diversity of predictions is a testament to our approach to better capturing such a one-to-many relationship. From an experimental point of view, this is useful as experimentalists may prefer a diverse set of synthesis recipes instead of a few (possibly due to the availability of compounds in the lab, safety, or apparatus constraints).

\paragraph{Generalization to new precursors.}
In retrosynthesis for materials discovery, experimental material scientists start from a target compound to identify potential precursor candidates that can be reacted to synthesize the desired target. This process involves screening extensive databases to find suitable compounds.
As a case study, consider the target compound \ce{Cr2AlB2}. A search in the Materials Project \citep{jain2013commentary} database yields potential precursor compounds such as \ce{B}, \ce{Cr}, \ce{Al}, \ce{AlB2}, \ce{Al45Cr7}, \ce{Al8Cr5}, \ce{AlCr2}, \ce{CrB}, \ce{CrB4}, \ce{Cr3B4}, \ce{Cr5B3}, \ce{Cr2B3}, \ce{CrB2}, \ce{Cr2B}, and \ce{Cr4B}. We train our model on a training set where none of these compounds were present, a scenario in which all previous methods would be \textit{inappliacble}. Inputting these into our model yields the reported synthesis route of reacting $\ce{CrB} + \ce{Al} \rightarrow \ce{Cr2AlB2}$ as the third highest-ranked synthesis recipe. \cref{table:new_prec} shows four more examples of reactions where Retro-\mbox{Rank-In} is able to correctly predict as its top-ranked precursor set, while Retrieval-Retro falls short (predicts $\emptyset$) due to it being limited to precursors seen during training. This underlines that Retro-\mbox{Rank-In} can lead to new precursor candidates, paving the way for more novel, potential synthesis pathways for target compounds. 

\begin{table}[ht]
    \label{table:new_prec}
    \centering
    \renewcommand{\arraystretch}{1.3}
    \setlength{\tabcolsep}{6pt}
    \caption{\textbf{Extrapolation to new precursors}: Examples from the NMS dataset where the ground truth precursors have not been part of the training set. Retrieval-Retro fails to predict the correct precursor candidates due to its inability to choose new precursors, while Retro-\mbox{Rank-In} correctly predicts the ground truth as the top-ranked precursor set.}
    \Large 
    \resizebox{\linewidth}{!}{
    \begin{tabular}{l  c  c  c}
        \toprule
        \textbf{Target material} & \textbf{Ground truth} & \textbf{Retro-Rank-In (ours)} & \textbf{Retrieval-Retro} \\
        \midrule
        Ba(GaSb)$_2$ & \{GaSb, Ba\} & \{GaSb, Ba\} & $\emptyset$ \\
        Na$_5$NpO$_6$ & \{Na$_2$CO$_3$, NpO$_2$\} & \{Na$_2$CO$_3$, NpO$_2$\} & $\emptyset$ \\
        AlBMo & \{BMo, Al\} & \{BMo, Al\} & $\emptyset$ \\
        Ga$_2$Mo$_2$C & \{Ga, Mo$_2$C\} & \{Ga, Mo$_2$C\} & $\emptyset$ \\
        \bottomrule
    \end{tabular}}
\end{table}

\paragraph{Suitability of sequence models.} 
We extend our method to select precursors sequentially from a given set. Specifically, we build the precursor set \(S\) by adding one precursor at a time, allowing the model to learn conditional probabilities of co-occurrences $p(\mathbf{x}_P|\mathbf{x}_T)$. However, when evaluated on the Distinct Reaction dataset, this approach shows limited competitiveness, yielding exact match accuracies of 18.71\%, 34.72\%, 39.83\%, and 43.78\% for Top-1, Top-3, Top-5, and Top-10 predictions, respectively. We attribute these modest results largely to the infrequency of strongly correlated precursor pairs (\cref{fig:prec_pairs_correlation}), which limits the model’s ability to learn meaningful joint distributions. Moreover, the necessity for a larger model to process the context of precursors already assigned to the set presents significant challenges when training in the low-data regime of our task.

\subsection{Model analysis}
\paragraph{Learning a pairwise ranker.} 
Moreover, we examine how reformulating the multi-label classification problem into learning a pairwise ranking influences the model's performance. For this analysis, we select MTEncoder embeddings. \cref{tab:ablation_graph} depicts the results of those combinations. Notably, the variants that learn the pairwise ranking demonstrate a significant performance enhancement over alternatives.

\begin{table}[htb!]
\centering
\caption{\textbf{Ablation on learning problem formulation.} We evaluate our approach using combinations of embeddings from MTEncoder, applied to either the pairwise ranker learning problem or multi-label classification. Bold for best.} 
\label{tab:ablation_graph}
\small 
\begin{tabular}
{lc@{\hspace{0.3cm}}c@{\hspace{0.3cm}}c@{\hspace{0.3cm}}c@{\hspace{0.3cm}}c}
\toprule
\multirow{2}{*}{\textbf{Embedding}} & \multirow{2}{*}{\textbf{Pairwise}} & \multicolumn{4}{c}{\textbf{Top-K Accuracy $\uparrow$}} \\ 
\cmidrule(l){3-6} 
 &  & Top-1 & Top-3 & Top-5 & Top-10 \\ 
\midrule
Composition      &    \textcolor{red}{\xmark }    & \textbf{42.31}  & 55.35 &59.36 &62.86  \\
& &\scriptsize (1.15) & \scriptsize (1.00) & \scriptsize (0.95) & \scriptsize (0.80)\\
Composition    & \textcolor{blue}{\checkmark} & 39.30 & \textbf{55.86}  & \textbf{63.09} &\textbf{71.56}  \\
& &\scriptsize (4.08) & \scriptsize (2.54) & \scriptsize (2.82) & \scriptsize (2.35)\\
\midrule
MTEncoder      &    \textcolor{red}{\xmark }    & \textbf{49.01}  & 64.59 & 68.78 &71.24  \\
& &\scriptsize (0.54) & \scriptsize (0.29) & \scriptsize (0.40) & \scriptsize (0.51)\\
MTEncoder    & \textcolor{blue}{\checkmark} & 48.93 & \textbf{65.45}  & \textbf{72.51 } & \textbf{78.48}  \\
& &\scriptsize (0.50) & \scriptsize (0.31) & \scriptsize (0.69) & \scriptsize (0.82)\\

\bottomrule
\end{tabular}
\normalsize 
\end{table}

\paragraph{Hard negative mining.}
In our experiments, we integrate hard negative mining into the training process to evaluate its impact on model performance. At each epoch, we increase the sampling probabilities of negative samples with high cosine similarity to the ground truth precursor set. Contrary to findings in other domains where hard negative mining enhances model accuracy \cite{moreira2024nv, monath2023improving}, our results indicated that this technique did not improve performance in our specific application.

\paragraph{Choice of information contextualization.} From \cref{tab:ablation_table_3}, all of the ablation variants (self-attention, concatenation, addpooling, transformer, and meanpooling) perform within a fairly narrow range—the Top‐1 through Top‐10 accuracy numbers are all close, and the standard deviations also suggest there the absence of a clear superior strategy. Because the performance differences are small, the simplest method becomes the most appealing choice in practice (Occam's razor \citep{domingos1999-occamsrazor}), as they provide results on par with the more complex approaches while being easier to implement and faster to train.

\textbf{Choice of ranker architectures.} As shown in \cref{tab:ablation_table_3}, all variants (self-attention, concatenation, addpool, transformer, meanpool) perform similarly, with close Top-1 to Top-10 accuracy and overlapping standard deviations. Given these minimal differences, the simplest method is preferable for efficiency and ease of implementation (Occam's razor \citep{domingos1999-occamsrazor}).

\begin{table}[tb]
\centering
\caption{\textbf{Ablation for ranker architecure.} We compare different ranker architectures.}
\label{tab:ablation_table_3}
\small 
\begin{tabular}
{l@{\hspace{0.6cm}}c@{\hspace{0.6cm}}c@{\hspace{0.6cm}}c@{\hspace{0.55cm}}c}
\toprule
\multirow{2}{*}{\textbf{Convolution}} & \multicolumn{4}{c}{\textbf{Top-K Accuracy $\uparrow$}} \\ 
\cmidrule(l){2-5} 
 & Top-1 & Top-3 & Top-5 & Top-10 \\ 
\midrule
Self-Attention             & 47.30 & 62.48  & 69.56 & \underline{75.99} \\
& \scriptsize (0.79) & \scriptsize (1.01) & \scriptsize (1.20) &  \scriptsize (1.10)\\
Concatenation             & 47.04 & 62.64  & \textbf{70.05} & \textbf{76.61}  \\
& \scriptsize (1.76) & \scriptsize (1.73) & \scriptsize (2.13) & \scriptsize (1.66)\\
Addpooling             & \textbf{48.48} & \textbf{63.36} & \underline{69.96} &75.85  \\
& \scriptsize (1.48) & \scriptsize (0.56) & \scriptsize (0.44) & \scriptsize (0.58)\\
Transformer             & 45.95 & 62.07 & 68.55 &74.65  \\
& \scriptsize (1.92) & \scriptsize (1.56) & \scriptsize (1.42) & \scriptsize (1.42)\\
Meanpooling & \underline{47.78} & \underline{63.13} & 69.35 & 75.78  \\
& \scriptsize (1.06) & \scriptsize (1.02) & \scriptsize (1.28) & \scriptsize (1.55)\\
\bottomrule

\end{tabular}
\normalsize 
\end{table}

\section{Concluding Remarks}

\paragraph{Limitations.}
While our approach represents a significant advancement, enabling the applications to novel synthesis routes and achieving state-of-the-art performance, it also has several limitations. Key synthesis parameters such as temperature, duration, and pressure, which are critical to determining the final synthesized materials, are not explicitly modeled \citep{huo2022machine}. Additionally, precursor interactions can result in the formation of intermediate compounds, which our current method does not fully account for due to the limited availability and documentation of such intermediates.  
Moreover, incorporating crystallographic structure data could further enhance predictive performance. However, the scarcity of datasets that integrate reaction pathways with structural information presents a challenge. Despite this, our approach is designed to be extensible and can incorporate such data when it becomes available. Resources like the Inorganic Crystal Structure Database (ICSD) \citep{zagorac2019recent} provide extensive collections of crystal structures, which could facilitate future improvements.

\paragraph{Summary.}
In this work, we introduced Retro-\mbox{Rank-In}, a novel ranking-based framework for inorganic retrosynthesis planning that implicitly incorporates broad chemical domain knowledge. Our approach redefines precursor prediction by learning a pairwise ranker that generalizes beyond known precursors, overcoming prior limitations and enabling the discovery of completely novel synthesis recipes. Comparative evaluations show that Retro-\mbox{Rank-In} sets the new state-of-the-art, particularly excelling in out-of-distribution scenarios. 
We will release the code upon acceptance to enable efficient synthesis planning throughout research labs.

\paragraph{Future Work.}
We identify several promising directions for future research.  
First, integrating structural data into precursor ranking could enhance prediction accuracy by better capturing crystallographic similarities. Larger pretrained models \cite{liao2022equiformer, neumann2024orb} could further refine structural understanding, incorporating domain knowledge crucial for precursor selection. Additionally, modeling the precursor ranking as a direct ranking between a target and two precursor candidates could explicitly enable the model to choose between precursors, improving interpretability and decision-making. 
Finally, analyzing attention patterns and the learned chemical space could provide insights into how the model captures chemical compatibility, revealing implicit reaction rules.

\newpage

\section*{Impact Statement}

This work aims to advance the field of inorganic retrosynthesis by improving the generalizability and predictive power of ML approaches. By providing a robust framework for retrosynthetic pathway prediction, Retro-\mbox{Rank-In} has the potential to accelerate the discovery of novel inorganic materials, which could lead to advancements in fields such as clean energy, catalysis, and electronic materials. While our method primarily serves scientific and industrial research, care should be taken to ensure that it is not misused for the synthesis of hazardous or environmentally harmful compounds. Future work should consider integrating ethical guidelines and safety constraints to mitigate potential risks associated with novel material discovery.

\section*{Acknowledgement}
The authors J.L.M.R. and T.P. wish to express their sincere gratitude to the Bavarian Ministry of Economic Affairs, Regional Development and Energy and TUMint.Energy Research GmbH for their generous support.

\section*{Author Contributions}
T.P. conceived the idea of a pairwise ranking-based approach, implemented the proposed methods, conducted experiments, and supervised the project from conception to analysis of results. A.T. contributed to project conception, supervision, and the idea of conditional generation, as well as manuscript conception and writing. E.P. was involved in project conception, result analysis, and manuscript writing. S.H. and M.L. prototyped initial components of our approach, implemented baseline models, analyzed their results,  and contributed to writing the main sections and composing figures. J.J. worked on baselines, explored alternative solutions for our approach, contributed to result analysis and visualization, and wrote parts of the supplementary information. T.W. was involved in results analysis, figure generation, early-stage conceptualization of our approach, and contributed to the supplementary information. M.P.F. assisted in implementing parts of the baselines, writing parts of the supplementary material and contributed to figure design and plotting. C.M. implemented the Mistral-based baseline. E.O. and J.L.M.R. offered insightful feedback on experimental design and analysis and on the manuscript. Their expertise in materials science was critical in refining the methodological approach.

\bibliography{references}
\bibliographystyle{icml2025}

\newpage

\input{appendix}

\end{document}

%% file: appendix.tex
\appendix
\onecolumn

\section{Notation} \label{app:notation}

\begin{table}[htb]
\centering
\caption{Mathematical notation overview.}
\label{tab:notation}
\begin{tabular}{lll}
\toprule

Symbol & Domain & Definition \\
\hline
\multicolumn{3}{l}{\textbf{General}} \\
$\mathbf{x} = (x_1, \dots, x_d)$ & $\mathbb{R}^d$ & Material composition \\
$x_i$ & $\mathbb{R}$ & Stochiometric fraction of element $i$ \\
$\tilde{\mathbf{x}}$ & $\mathbb{R}^{h}$ & Material embedding \\
$\textit{P}$ & $\mathbb{R}^{d}$ & Precursor material \\
$\textit{S}$ & $\mathbb{R}^{m\times d}$ & Precursor set \\
$\textit{T}$ & $\mathbb{R}^{d}$ & Target material \\
$p$ & $\mathbb{P}$ & Probability of a precursor or set \\
$y$ & $\{0,1\}$ & Whether a target-precursor pair in dataset \\
$\theta$ & -- & Parameterized learned model \\
$\mathbb{B}$ & -- & Binary classifier \\
$\mathcal{L}$ & -- & Loss function \\
\hline
\multicolumn{3}{l}{\textbf{Dimensions}} \\
$d$ & $\mathbb{N}$ & Dimension of composition vector \\
$h$ & $\mathbb{N}$ & Hidden dimension \\
$m$ & $\mathbb{N}$ & Number of precursors per set \\
$n$ & $\mathbb{N}$ & Number of unique precursors per set \\
$N$ & $\mathbb{N}$ & Number of unique precursor in dataset \\
$K$ & $\mathbb{N}$ & Top-$K$ ranked precursor sets \\
\hline
\multicolumn{3}{l}{\textbf{MTEncoder}} \\
$\mathbf{e}$ & $\mathbb{R}^{h}$ & Learned chemical element embedding \\
$\mathbf{f}$ & $\mathbb{R}^{h}$ & Sinusoidal fractional embedding \\
$\mathbf{z}$ & $\mathbb{R}^{h}$ & Per-element embedding \\
$\mathbf{t}$ & $\mathbb{R}^{h}$ & Compound embedding \\
$\mathbf{s}$ & $\mathbb{R}^{(k+1) \times h}$ & MTE input sequence \\
$k$ & $\mathbb{N}$ & Number distinct elements in composition \\
\bottomrule
\end{tabular}
\end{table}

Example: 

\begin{equation}
    T \leftarrow \underbrace{\{P_1, P_2, P_3 \}}_{r=3} \equiv \{ \underbrace{\{A,B\}}_{m_1=2}, \{A,C,F\}, \{C,G\} \}
\end{equation}

\begin{itemize}
    \item $n=5$ (explanation: there are five unique precursors $\{A,B,C,F,G\}$)
    \item $N$ is a large dataset-dependent number.
\end{itemize}

\section{Dataset}

\begin{table}[h]
    \centering
    \caption{Dataset Statistics including Train, Validation, and Test Splits}
    \label{tab:dataset_stats}
    \begin{tabular}{lccc}
        \toprule
        Dataset & Train & Validation & Test \\
        \midrule
        Complete Reaction Archive    & 9715  & 2430       & 6659 \\
        Distinct Reactions  & 5091  & 1274       & 2893 \\
        Novel Materials Systems    & 3012  & 753        & 2892 \\
        \bottomrule
    \end{tabular}
\end{table}

\label{Appendix:Dataset}

\section{Implementation Details}
Our code will be made available on GitHub upon publication of the manuscript.



\subsection{MTEncoder}
\label{appendix_mte}
\begin{figure}[htb!]
    \centering
        \centering
        \includegraphics[width=0.35\columnwidth]{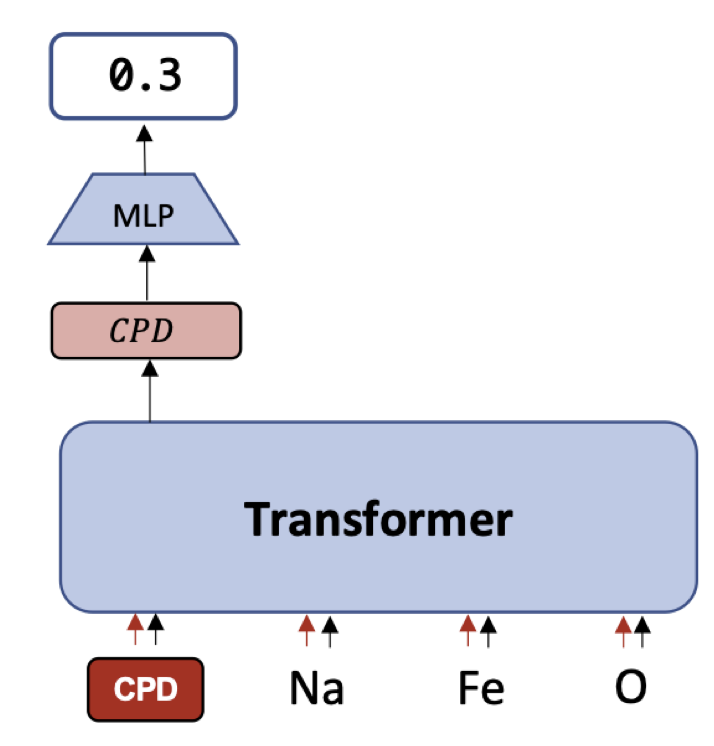}
        \caption{\textbf{MTEncoder architecture overview.}
        This diagram illustrates the MTEncoder framework, where material compositions are tokenized and processed through a transformer model.}
        \label{fig:plotmte}
\end{figure}

\cref{fig:plotmte} presents a schematic representation of the MTEncoder architecture, illustrating how material compositions are processed using a transformer-based encoder. The input consists of element tokens (Na, Fe, O), along with a Compound special token (\textit{CPD}), which aggregates information from the constituent elements. These inputs are passed through a transformer model, which learns a contextualized representation of the material composition. The CPD token serves as the learned materials representation, and is then fed into MLP classifiers for property prediction. 


\begin{table}[h]
    \centering
    \begin{tabular}{l}
        \toprule
        \textbf{Pretraining Tasks} \\
        \midrule
        Stress \\
        Band Gap (Direct) \\
        Band Gap (Indirect) \\
        DOS at Fermi Level \\
        Energy Above Hull \\
        Formation Energy \\
        Corrected Energy \\
        Phase Separation Energy \\
        Number of Sites \\
        Total Magnetization \\
        Space Group \\
        Masked Element Modelling (Self-supervised) \\
        \bottomrule
    \end{tabular}
    \caption{\textbf{Pretraining tasks}
    Tasks during MTEncoder pretraining, data is used from alexandria.}
    \label{tab:pretraining_tasks}
\end{table}

\subsection{Training Details}

\begin{figure}[htb!]
    \label{figure_layer_ablation}
    \centering
        \centering
        \includegraphics[width=0.6\columnwidth]{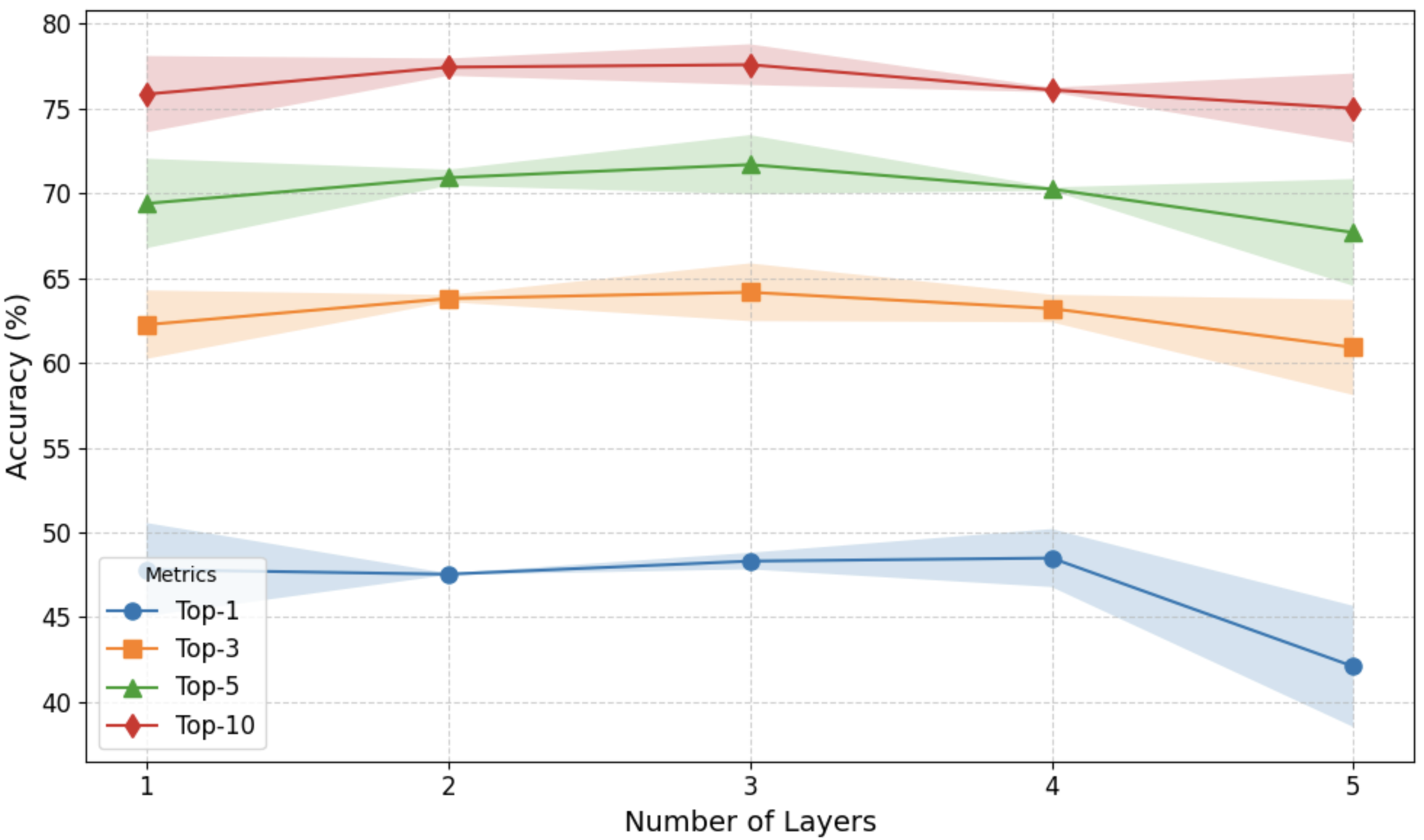}
        \caption{\textbf{Ablation for layers.} Retro-Rank-In tested for various numbers of layers. Results show the robustness of our method regarding hyperparameter choice.}
        \label{fig:layers}
\end{figure}

\paragraph{Model Training.}
We ablate over the number of layers in Retro-Rank-In to assess robustness against this hyperparameter. As shown in \cref{fig:layers}, the model performs consistently well under various depths, peaking around three layers for most metrics. Top-1 accuracy dips slightly at both extremes (one and five layers) but remains stable near the center. Similar trends hold for the other metrics (Top-3, Top-5, Top-10), suggesting that while some tuning of depth may help refine results, the method is generally resilient to layer variations.

\paragraph{Hyperparameters}
For our model, we conducted further hyperparameter tuning with ranges specified in Table \ref{tab:hyperparamter_table}. We explored the following parameter spaces: batch size B in \{128, 256, 512\}, number of attention heads H in \{1, 4, 8\}, number of feedforward (FFWD) layers L in \{1,2,3,4\}, learning rate $\eta$ in [$10^{-5}$, $10^{-3}$] and MTEncoder learning rate $\eta_{MT}$ in [$10^{-6}$, $10^{-4}$]. The optimal configuration was determined based on model performance on the validation set, resulting in the following selected values: batch size of 128, 1 attention head, 3 FFWD layers, learning rate of $6.81 \times 10^{-5}$ and MTEncoder learning rate of $6.37 \times 10^{-5}$. We report test performance using these optimized parameters.
\begin{table}[htb]
\centering
\caption{Hyperparameter configuration of Retro-Rank-In}
\label{tab:hyperparamter_table}
\begin{tabular}
{l@{\hspace{0.6cm}}c@{\hspace{0.6cm}}c@{\hspace{0.6cm}}c@{\hspace{0.55cm}}c}
\toprule
\multirow{2}{*}{\textbf{Hyperparameters}} & \multicolumn{2}{c}{\textbf{Configuration}} \\ 
\cmidrule(l){2-3} 
 & Search Space & Selected Values \\ 
\midrule
Batch Size (B)& \{128, 256, 512\} & 128 \\
Attention Heads (H)& \{1, 4, 8\} & 1 \\
FFWD Layers (L)& \{1,2,3,4\} & 3 \\
Learning Rate ($\eta$) & [$10^{-5}, 10^{-3}$] & $6.81 \times 10^{-5}$ \\
MT Learning Rate ($\eta_{MT}$) & [$10^{-6}, 10^{-4}$] & $6.37 \times 10^{-5}$ \\
\bottomrule

\end{tabular}
\end{table}

\section{Baseline Methods} 

\subsection{Retrieval-Retro}
Retrieval-Retro~\cite{noh2024retrieval} proposes a two-stage approach to inorganic retrosynthesis that implicitly extracts the precursor information of reference materials. First, for each target material, reference materials from the knowledge base of previously synthesized materials are elaborately retrieved by two complementary models: Inspired by Synthesis Similarity~\cite{he2023precursor}, the MPC retriever, trained for Masked Precursor Completion (MPC), selects reference materials sharing similar precursors. The Neural Reaction Energy (NRE) Retriever integrates domain knowledge and leverages the thermodynamic relationships between materials to identify precursor sets with a high probability of synthesizing the target. Representing target and reference materials as fully connected composition graphs, the final Retrieval-Retro stage then employs self-attention and cross-attention mechanisms to implicitly extract relevant precursor information from the reference materials and predict precursor sets based on the probability for each individual precursor.

\subsection{ElemwiseRetro}
ElemwiseRetro~\cite{kim2022element} proposes a template-based approach to inorganic retrosynthesis that represents target materials as fully connected composition graphs. To guide the retrosynthesis process, the researchers distinguish between two types of elements: "source elements," provided as precursors, and "non-source elements," which either appear or disappear during the reaction. For each source element in a target composition, their model predicts the most likely anionic framework—a composition of non-source elements—from a predefined set of templates. The selected source element and its template are then concatenated to form the actual precursor compound, which may be reformulated using a stoichiometric lookup table to ensure frequent and chemically valid compositions.

\subsection{Synthesis Similarity}
\citet{he2023precursor} uses a similarity-based approach to identify precursor sets for inorganic retrosynthesis. They introduce a vector representation for Masked Precursor Completion (MPC) and chemical composition recovery tasks and use this encoding to retrieve reference materials similar to a given target. They initialize the prediction with the precursor set of the reference material and use the MPC network to complement the prediction for a valid precursor set.

\subsection{Mistral 7B}
We employ the Mistral 7B model for LM-based precursor prediction, initially experimenting with few-shot prompting. However, due to insufficient performance, we shifted to a more structured prompting approach. The prompt was as follows: \begin{quote} "You are tasked with identifying precursors for synthesizing the target material \( \text{Mn}_{0.71}\text{Zn}_{0.21}\text{Fe}_{2.08}\text{O}_4 \). You should choose exactly 3 precursors. Generate 20 possible precursor material combinations in descending order of probability. Each route should represent a unique combination of precursors likely to result in the target material. Use the chemical formulas for all precursors instead of their common names. Output them in a Python list format, where each precursor is a string, and each possible combination is a list. Each list should have a length of 3. The response should only include a list of lists where the smaller the index of the list, the higher the probability that the precursor combination has. Do not add any other sentences to the response, only print the list."
\end{quote}This structured prompt, directing the model to rank precursor combinations by probability, improved accuracy. Post-processing, including element validation and duplicate removal, further refined results, making the structured approach more efficient than few-shot prompting

\begin{table*}[htb!]
\centering
\small 
\caption{\textbf{Performance results for Mistral.} Mistral evaluated across three datasets: (a) Complete Reaction Archive, (b) Distinct Reactions, and (c) Novel Material Systems. Bold values indicate the best performance and underline the second best. All scores are reported as averages over five runs, with standard deviations in parentheses.}
\resizebox{\linewidth}{!}{
\begin{tabular}
{lcccccccccccccc}
\toprule
 & \multicolumn{4}{c}{\textbf{(a) Complete Reaction Archive}} & \multicolumn{6}{c}{\textbf{(b) Distinct Reactions}} & \multicolumn{4}{c}{\textbf{(c) Novel Materials Systems}} \\
\cmidrule(lr){2-5} \cmidrule(lr){7-10} \cmidrule(l){12-15}
\textbf{Model}      & \multicolumn{4}{c}{\textbf{Top-K Accuracy $\uparrow$}} & \multicolumn{6}{c}{\textbf{Top-K Accuracy $\uparrow$}} & \multicolumn{4}{c}{\textbf{Top-K Accuracy $\uparrow$}} \\
\cmidrule(lr){2-5} \cmidrule(lr){7-10} \cmidrule(l){12-15}
      & Top-1 & Top-3 & Top-5 & Top-10 & & 
      Top-1 & Top-3 & Top-5 & Top-10 & &
      Top-1 & Top-3 & Top-5 & Top-10 \\
\midrule
Mistral 7B & 24.75 & 35.34 & 37.62 & 39.78 & & 16.91 & 22.30 & 24.27 & 25.93 & & 16.91 & 22.30 & 24.27 & 25.93 \\
\bottomrule
\end{tabular}}
\end{table*}

\section{Evaluation Protocol}
\label{appendix:eval}

To mitigate the combinatorial explosion, we first select the $30$ precursor candidates with the highest probabilities from the model's predictions. \cref{tab:top-k-analysis} illustrates how increasing this sample size improves Top-K match accuracy but also amplifies computational complexity, as more precursor combinations must be evaluated. These selected precursors are then combined to form candidate sets, and a set is deemed valid if the union of its elements contains all elements of the target composition. Each valid set is assigned a joint probability derived from the probabilities of its individual precursors, and the valid sets are subsequently ranked in descending order by this joint probability. For the Top-K match accuracy, we focus on the subset of $K$ highest-ranked valid sets. If the ground-truth precursor set $S_{\text{true}}$ matches any of these K sets, we assign a score of 1 for that target; otherwise, we assign 0. Finally, we compute the overall score by averaging these individual scores across the entire test set.

We evaluate all approaches in this way, except for Mistral and ElemwiseRetro, which already output the precursor sets. Therefore, we skip the step of constructing the sets from the candidates. The remaining part of the evaluation stays the same.

\vspace{-5mm}
\begin{table}[htb]
\centering
\caption{\textbf{Top-K Match Accuracy across candidate precursor sample sizes $n$.} Investigating the effect of varying amounts of top $n$ candidate precursors sampled based on highest probabilities (see \cref{appendix:eval})), on example Top-K accuracy of Retro-Rank-In, and on the total number of evaluated precursor combinations $N$.}
\label{tab:top-k-analysis}
\small 
\begin{tabular}
{l@{\hspace{0.6cm}}c@{\hspace{0.6cm}}c@{\hspace{0.6cm}}c@{\hspace{0.6cm}}c@{\hspace{0.6cm}}c}
\toprule
\multirow{2}{*}{\textbf{$n$}} & \multicolumn{4}{c}{\textbf{Top-K Accuracy $\uparrow$}} & \multirow{2}{*}{\textbf{$N$}} \\ 
\cmidrule(l){2-5} 
 & Top-1 & Top-3 & Top-5 & Top-10\\ 
\midrule
10 & 43.22 & 57.05  & 62.34 & 67.95 & 357,416\\
20 & 43.26 & 57.26  & 62.59 & 68.78 & 7,333,299\\
{\textbf{30}} & 43.30 & 57.43  & 62.90 & 69.29 & 58,386,823\\
40 & 43.57 & 57.81  & 63.31 & 69.81 & 320,996,717\\
\bottomrule

\end{tabular}
\normalsize 
\end{table}

\section{Further Analysis}

\subsection{Hyperparameter ablation} 
We examined the impact of network depth on our model's performance by training Retro-Rank-In with feedforward layers ranging from 1 to 5. With each additional layer, we reduced the dimensionality by a factor of two. Our findings indicate that the model is robust to these hyperparameter variations, with a three-layer architecture yielding the highest Top-1, Top-3, and Top-5 accuracy scores. Consequently, we selected this as our default configuration (\cref{figure_layer_ablation}).

\subsection{Pretraining encoder ablation} 
\cref{tab:ablation_table_3} further examines the impact of no pretraining versus pretraining for the encoder model. We observe the Top-K accuracy drastically decrease without having a pretrained encoder.

\begin{table}[H]
\centering
\caption{\textbf{Ablation pretraining.} Investigating the impact of pretraining for the encoder with the Top-K accuracy.}
\label{tab:ablation_table_3}
\small 
\resizebox{0.5\linewidth}{!}{
\begin{tabular}
{l@{\hspace{0.6cm}}c@{\hspace{0.6cm}}c@{\hspace{0.6cm}}c@{\hspace{0.55cm}}c}
\toprule
\multirow{2}{*}{\textbf{Pretrained Encoder}} & \multicolumn{4}{c}{\textbf{Top-K Accuracy $\uparrow$}} \\ 
\cmidrule(l){2-5} 
 & Top-1 & Top-3 & Top-5 & Top-10 \\ 
\midrule
\textcolor{red}{\xmark }            &  33.24 & 53.13 & 62.70 & 71.22 \\
& \scriptsize (8.20) & \scriptsize (5.52) & \scriptsize (5.45) & \scriptsize (2.98)\\
\textcolor{blue}{\checkmark }           & \textbf{47.04} & \textbf{62.64} & \textbf{70.05} & \textbf{76.61}  \\
& \scriptsize (1.76) & \scriptsize (1.73) & \scriptsize (2.13) & \scriptsize (1.66)\\

\bottomrule
\end{tabular}}
\normalsize 
\end{table}

\subsection{Performance across different chemistries}

\begin{figure}
    \centering
        \centering
        \includegraphics[width=0.55\columnwidth]{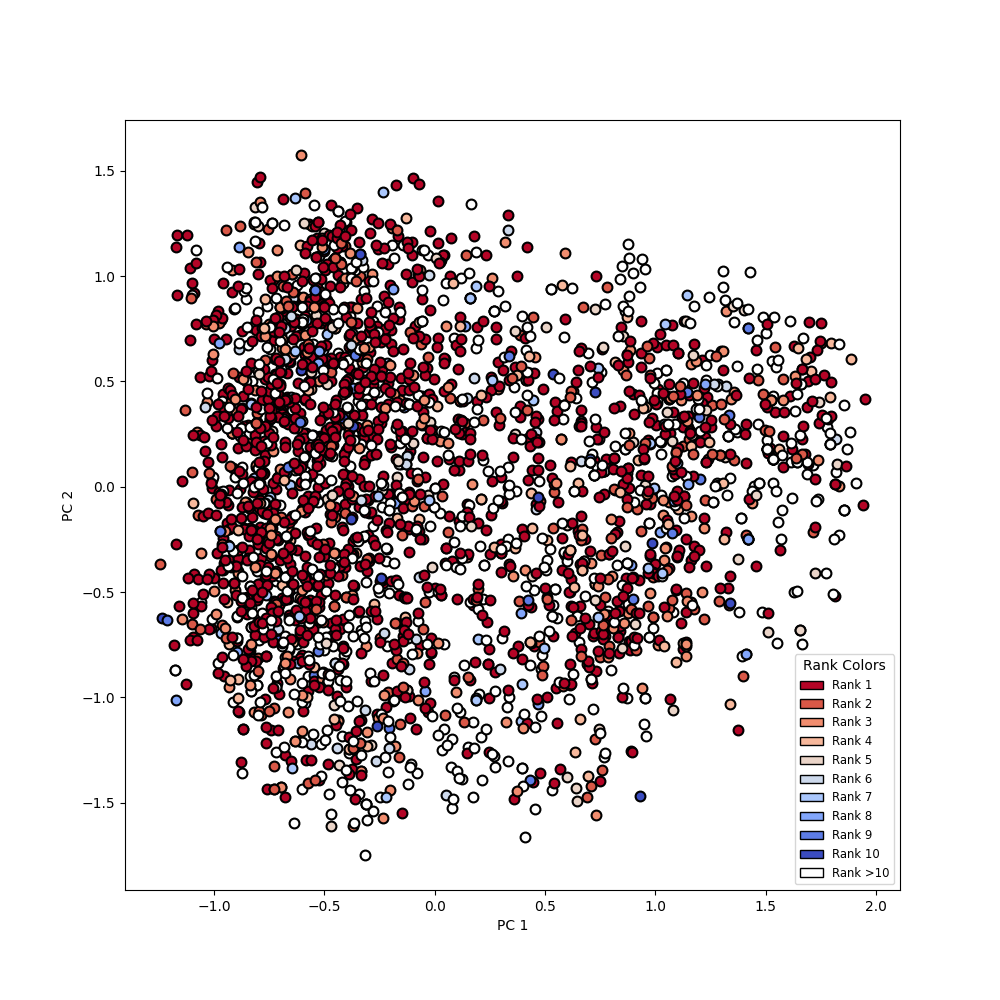}
        \caption{\textbf{PCA Visualization of target embeddings with rank-based coloring.} 
        PCA of target embeddings, where each point is color-coded based on rank. Higher-ranked points are shown in warmer tones, while lower-ranked ones appear in cooler shades, illustrating the distribution of rankings in the embedding space.}
        \label{tab:pca_dec}
\end{figure}

In Figure~\ref{tab:pca_dec}, we illustrate the correlation between the target embeddings and the ranks assigned by our model, namely the Retro-Rank-In. To achieve this, we first process the chemical composition of each target material using the \texttt{Composition} class from the \texttt{pymatgen} library, which parses the input material string and standardizes its representation to ensure consistency in the interpretation of the chemical formula. The standardized composition was then encoded using the MTEncoder model \cite{prein2023mtencoder}, which maps the material string to a $h$-dimensional tensor representation. In this study, an embedding dimension of $h=512$ was used to capture the essential features of each material for further computational analysis. After acquiring the MTE embeddings, we projected them into a 2D space using Principal Component Analysis (PCA), where each point's color represents its assigned rank. Warmer hues (red) represent higher ranks, while cooler tones (blue to white) indicate lower ranks. The predominance of red points on this plane indicates that most embeddings were correctly classified as Rank 1, demonstrating strong model performance. In contrast, lower-ranking embeddings (e.g., those in blue) appear sparser, occupying smaller regions of the plot. This suggests that while the majority of materials achieve Rank 1, fewer are assigned to lower ranks. In summary, this two‐dimensional projection thus highlights the distribution of performance across the embedding space, revealing that the majority of embeddings cluster at higher ranks while relatively few reside in the lower‐rank region.

\subsection{Precusor correlation}

\begin{figure}[htb!]
    \centering
        \centering
        \includegraphics[width=0.45\columnwidth]{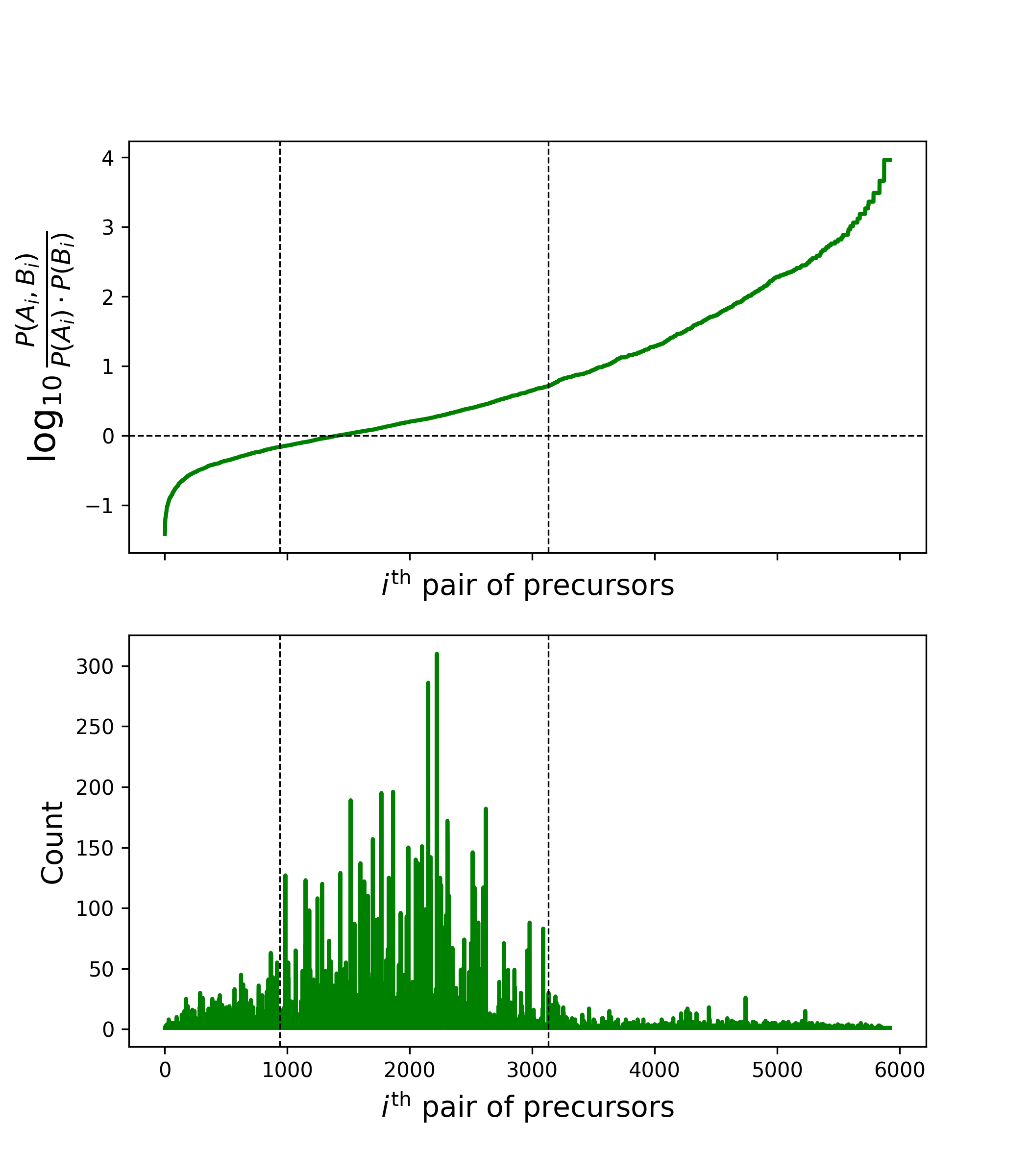}
        \caption{\textbf{Precursor pair correlation.} 
        The plot illustrates the correlation between pairs of precursors. Each point corresponds to a unique precursor pair, sorted along the x-axis by the strength of their correlation. Correlation is quantified here by the logarithm of the ratio of their joint probability to the product of their individual probabilities.}
 \label{fig:prec_pairs_correlation}
\end{figure}

The top plot of \cref{fig:prec_pairs_correlation} is a significant degree of positive and negative correlation for a fair amount of precursor pairs. The frequency of occurrence for the same pairs is visualized in the bottom figure. While a few unique precursor pairs show strong positive or negative correlations, the vast majority of frequently occurring pairs exhibit little to no correlation.

\subsection{Prediction diversity}

\begin{figure}
    \centering
        \centering
        \includegraphics[width=0.4\columnwidth]{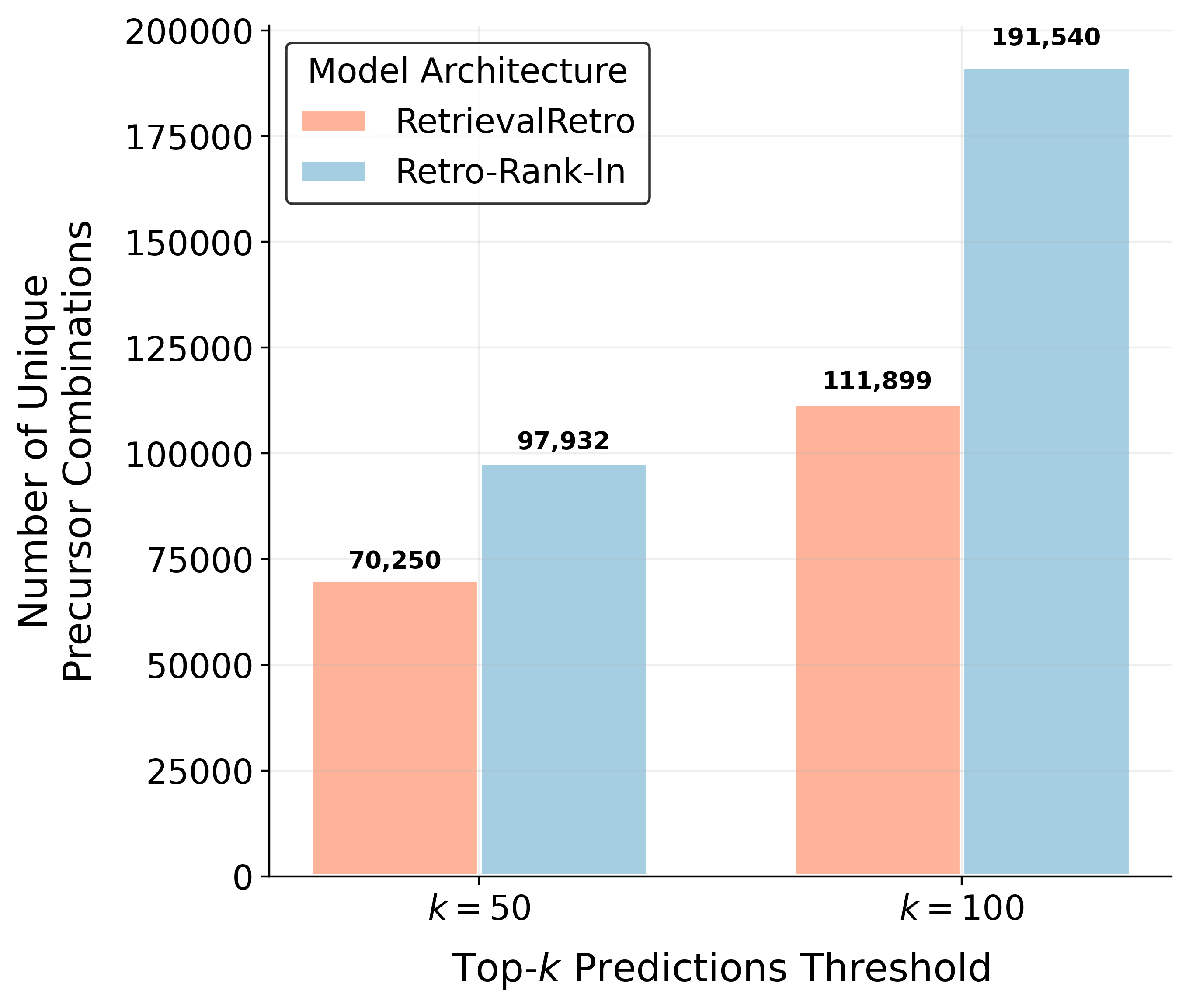}
        \caption{\textbf{Precursor set diversity.} Comparison of the number of unique precursor combinations generated by Retrieval-Retro and Retro-Rank-In.}
        \label{fig:plot10}
\end{figure}

\cref{fig:plot10} compares the number of unique precursor combinations identified by the two models, RetrievalRetro and Retro-Rank-In, at Top-K thresholds of K=50 and K=100. We observe that Retro-Rank-In consistently generates a greater number of unique precursor combinations than RetrievalRetro. This result suggests that Retro-Rank-In can capture a broader space of potential synthetic routes, which is advantageous for identifying novel and efficient strategies in retrosynthetic planning.

\begin{figure}
    \centering
        \centering
        \includegraphics[width=0.9\columnwidth]{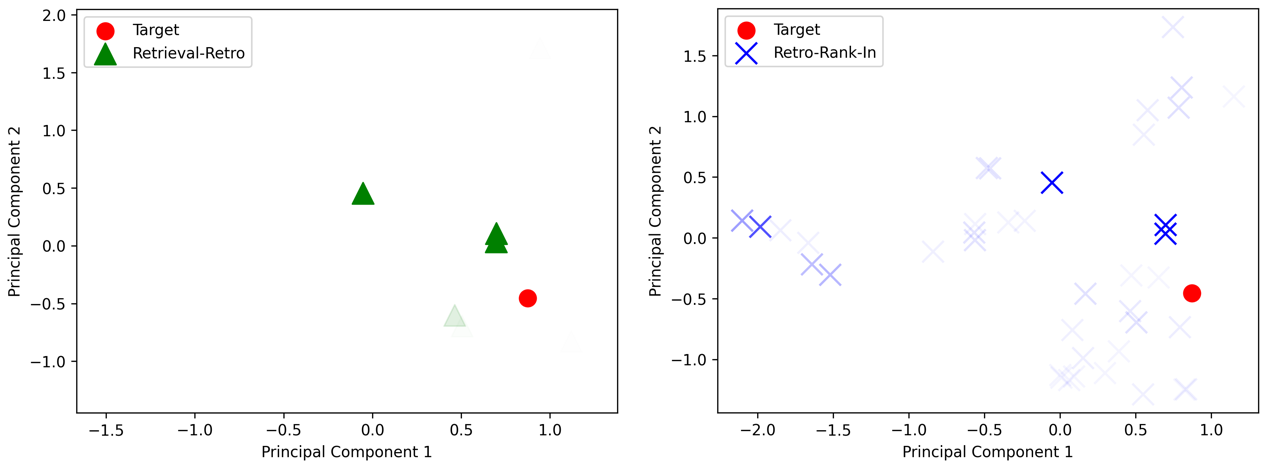}
        \caption{\textbf{Retro-Rank-In achieves higher diversity of predicted precursors.} PCA plot of a MTEncoder-encoded target material (red color) and precursors predicted by Retrieval-Retro (green triangles, left) and Retro-Rank-In (blue crosses, right). The intensity/alpha of each point is proportional to the probability assigned by each model. Clearly, Retrieval-Retro assigns high probabilities to a small number of precursors, leading to low diversity. In contrast, Retro-Rank-In assigns significant probabilities to a higher number of precursors, leading to higher diversity. Importantly, this improvement in diversity does not come at the expense of accuracy, as shown in Table \ref{table1:performance_comparison}.
        }
        \label{fig:plot9}
\end{figure}

This is further supported by the findings of \cref{fig:plot9}, which shows how each model allocates probability mass across its proposed precursor sets. In particular, RetrievalRetro tends to concentrate most of its probability on just a few highly ranked precursor combinations, as evidenced by a steep probability drop‐off after its top‐ranked suggestions. By contrast, Retro‐Rank‐In spreads its probability more evenly across a larger set of potential combinations, indicating a more diverse exploration of the synthetic space. Crucially, this broader coverage means Retro‐Rank‐In is less likely to overlook innovative or less obvious routes during retrosynthetic planning, offering a more comprehensive foundation for subsequent experimental validation.

\begin{figure}[htb!]
    \centering
        \centering
        \includegraphics[width=0.6\columnwidth]{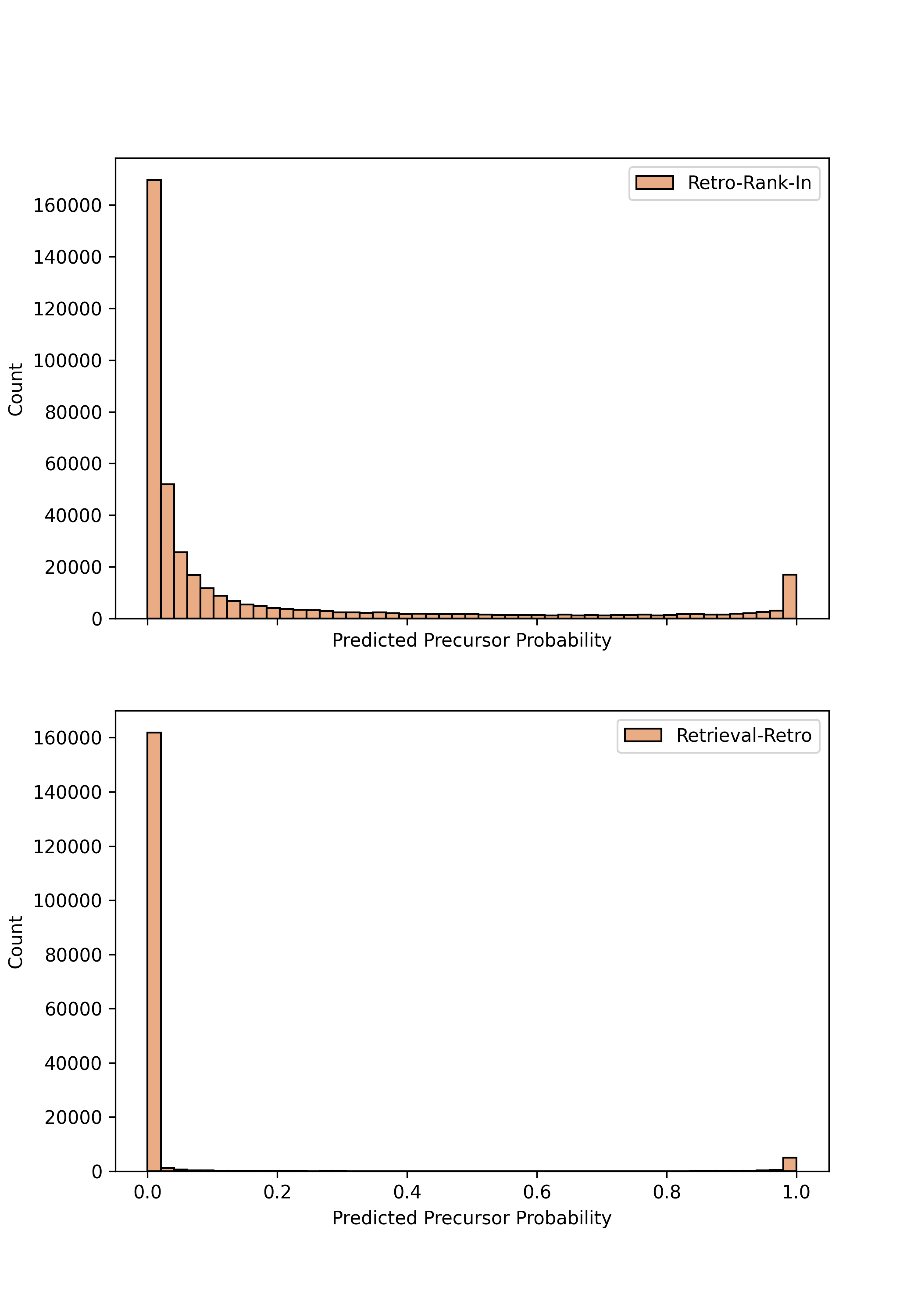}
        \caption{\textbf{Distribution of predicted probabilities.} A comparison of the top 60 highest predicted precursor probabilities across the test set of the Distinct Reactions dataset. Our approach demonstrates improved probability calibration compared to the previous state-of-the-art \citep{noh2024retrieval}. We attribute this improvement to the class-balanced learning of a pairwise ranker, which translates to enhanced throughout performance at higher values of $K$ in Top-$K$ exact match accuracy.}
        \label{fig:rr_ours_prob_dist}
\end{figure}

Lastly, \cref{fig:rr_ours_prob_dist} compares the distribution of predicted precursor probabilities (the top 60 highest values) for Retro-Rank-In (top) and Retrieval-Retro (bottom) on the Distinct Reactions test set. Retro-Rank-In produces a wider spread of mid-to-high probabilities, suggesting more nuanced confidence estimates, while Retrieval-Retro’s predictions cluster near zero or one. This pattern indicates that Retro-Rank-In is better calibrated, resulting in stronger performance at higher values of K in Top-K exact match accuracy.